\documentclass{iopart}
\usepackage{iopams}
\usepackage{graphicx}

\def\beq{\begin{equation}}
\def\eeq{\end{equation}}
\def\beqa{\begin{eqnarray}}
\def\eeqa{\end{eqnarray}}

\def\N{ {\cal N} }
\def\o{\mathrm{o}}
\def\f{\mathrm{f}}
\def\q{\mathbf{q}}
\def\r{\mathbf{r}}
\def\p{\mathbf{p}}
\def\j{\mathbf{j}}
\def\J{\hat{J}}

\def\v{\boldsymbol v}
\def\s{\boldsymbol s}
\def\u{\boldsymbol u}

\begin{document}

\title{The semiclassical continuity equation for open chaotic systems}

\author{Jack Kuipers, Daniel Waltner, Martha Guti\'errez and Klaus Richter}

\address{Institut f\"ur Theoretische Physik, Universit\"at Regensburg, D-93040 Regensburg, Germany}
\ead{Jack.Kuipers@physik.uni-regensburg.de}
\begin{abstract}
We consider the continuity equation for open chaotic quantum systems in the semiclassical limit.  First we explicitly calculate a semiclassical expansion for the probability current density using an expression based on classical trajectories.  The current density is related to the survival probability via the continuity equation, and we show that this relation is satisfied within the semiclassical approximation to all orders.  For this we develop recursion relation arguments which connect the trajectory structures involved for the survival probability, which travel from one point in the bulk to another, to those structures involved for the current density, which travel from the bulk to the lead.  The current density can also be linked, via another continuity equation, to a correlation function of the scattering matrix whose semiclassical approximation is expressed in terms of trajectories that start and end in the lead.  We also show that this continuity equation holds to all orders.
\end{abstract}


\section{Introduction} \label{intro}

The continuity equation is one of the basic concepts in physics, which appears in different contexts describing the conservation of a quantity.  For example in electromagnetism it expresses charge conservation, in fluid dynamics mass conservation, and in quantum mechanics it represents the  conservation of probability.  Quantum mechanically the continuity equation can be expressed as
\beq \label{QCE}
\frac{\partial \rho(\r,t)}{\partial t}+\nabla\cdot \j(\r,t)=0,
\eeq
where $\rho(\r,t)=|\psi(\r,t)|^2 $ is the probability density,
\beq \label{jrteqn}
\j(\r,t)= \frac{\hbar}{2m\rmi}[\psi^{*}(\r,t)\nabla \psi(\r,t)-\psi(\r,t)\nabla \psi^{*}(\r,t)],
\eeq
is the probability current density, and $\psi(\r,t)$ is the solution of the time dependent Schr\"odinger equation with initial condition $\psi_{\o}({\mathbf r})$ at $t=0$.  A classical analogue of this continuity equation is \cite{gaspard98}
\beq \label{CCE}
\frac{\partial C(\r,\r',t)}{\partial t}+\frac{1}{m}\nabla_{\r} \cdot \left(C(\r,\r',t)\, \p  \right)=0,
\eeq
where
\beq
C(\r,\r',t)= \det\left(-\frac{\partial^2 S(\r,\r',t)}{\partial \r \partial \r' }\right),
\eeq
is known as the Morette-Van-Hove determinant, $\p=\nabla_{\r}S (\r,\r',t)$ is the momentum at $\r$, and $S(\r,\r',t)=\int_{\r'}^{\r} \p \rmd \q-E t$ is the action principal function, satisfying the Hamilton-Jacobi equation \cite{arnold78}. A first connection between the two continuity equations can be made through the WKB ansatz \cite{bb03}, which corresponds to taking $\psi^{\rm WKB}({\mathbf r},t)= A ({\mathbf r},t){\rm e}^{\frac{i}{\hbar}S({\mathbf r},t)}$. By substituting this function into the Schr\"odinger equation, and neglecting higher order terms in $\hbar$ one arrives at the Hamilton-Jacobi equation, while the next order term in $\hbar$ leads to the classical continuity equation \eref{CCE}, upon the identification $\vert A\vert ^2=\vert C\vert$.

It is then straight-forward, from this ansatz in the WKB approximation, to preserve the conservation of probability at the quantum level for small $\hbar$, since this just relies on the classical preservation of probability. Moreover, this result is independent of the dynamics, and completely general for all Hamiltonian systems, though to our knowledge, there has not been much discussion about higher $\hbar$-dependent terms.  The semiclassical methods which we use in this paper have the advantage though of providing us with additional information, allowing us for example to explicitly calculate quantum corrections to the survival probability.  However, when applying these methods and semiclassical approximations, it is in fact an important issue to ensure that the unitarity of the quantum evolution is preserved, which in turn assures the fulfilment of \eref{QCE}.  Our aim in this paper is to show explicitly that this is indeed the case for chaotic systems.

In the context of transport through mesoscopic chaotic cavities \cite{richter00,jalabert00} this problem has been of wide interest. A possible way of describing the conductance of such a system is within the scattering approach \cite{datta95}.  The conductance is given in terms of asymptotic states in the leads projected onto Greens functions, which can be expressed semiclassically in terms of trajectories travelling from one lead to the other. The scattering matrix that describes the probability amplitudes of going from one channel to another, is unitary, which implies that the sum of all the probabilities should be equal to one.  It has been possible to recover this result semiclassically by considering interference terms due to correlated trajectories, which were first considered for periodic orbit correlations in the context of spectral statistics.  Moving beyond the diagonal approximation \cite{berry85}, the contribution of the first such correlated pair was found in \cite{sr01}, and this was later extended to all orders in \cite{muller04,muller05}.  These ideas were then applied to transport in calculating the conductance, with the first contribution calculated in \cite{rs02} and the extension to all orders in \cite{heusler06}.  For the scattering matrix, it follows that the average sum of the probabilities is indeed one, and in fact there are also no fluctuations around this value \cite{muller07,ks08}.

The continuity equation implies that the unitarity of the scattering matrix should be independent of the position of the cross-sections from which the scattering matrix is defined. Through this implication it is possible to show quantum mechanically that the scattering approach is equivalent \cite{jbs90,bjs93b} to the Kubo-linear response theory \cite{datta95},  where the conductance can be written in terms of states inside the scattering region.  It remains to be shown that this equivalence holds after applying semiclassical techniques.  However, a mechanism for relating trajectories in the bulk with escaping trajectories was discussed in \cite{waltner08} for the continuity equation itself. They studied the spatially integrated form of the continuity equation \eref{QCE}, which is given by
\beq
\label{intconteqn}
\frac{\partial }{\partial t}\rho (t)+\int _{S} \j(\r,t)\cdot \hat{n}_x \rmd x=0,
\eeq
where $S$ is the cross-section of the opening and $\hat{n}_x $ is the vector normal to this section at the point $x$ in $S$.  The survival probability $\rho (t)$ is given by the integral $\int_A \rmd \r \: \rho(\r,t)$ over the volume $A$ of the corresponding closed system.  To satisfy this continuity equation it was shown \cite{waltner08} that it is necessary to include `one-leg-loops', where the self-encounter now overlaps with the start or the end of the trajectory, among the correlated trajectories considered.  Also, and further in \cite{gutierrez08}, it was shown that `one-leg-loops' were a basic ingredient needed to correctly calculate the survival probability.  They are therefore necessary to recover unitarity, both of the survival probability and of the flow as expressed through the continuity equation.

In this article we wish to build on these and previous semiclassical works to show that the conservation of probability described by the continuity equation is preserved, to all orders, when we take the semiclassical approximation to the propagator and perform the semiclassical expansion of the survival probability and current density.  This approximation leads to expressions involving pairs of trajectories, and following the general philosophy of previous works we need to consider correlated pairs that can contribute in the semiclassical limit.  As shown in \cite{sr01,muller04,rs02,heusler06} these come from trajectories that have close self-encounters.  For chaotic systems, the property of local hyperbolicity means that we can construct a partner that differs in the encounter, leading to a small action difference and a contribution in the semiclassical limit, while the property of global ergodicity (and mixing) means that we can estimate the number of such pairs and calculate the contributions.  These two properties, and such pairs of trajectories, are therefore responsible for the universal behaviour exhibited by chaotic systems, and the reason why we consider such systems here.     

In particular, among the correlated trajectories which we consider in this article, we also include those contributions coming from pairs of trajectories involving `one-leg-loops' \cite{waltner08,gutierrez08}. Here we extend that work and show that the continuity equation is satisfied to all orders in the semiclassical approximation. For this purpose, we first re-examine the types of correlated trajectories which contribute and calculate a semiclassical expansion for the spatially integrated current density.  By integrating the current density with respect to time we can compare with the previous result for the survival probability \cite{gutierrez08}, to find agreement in line with the continuity equation.  We then derive recursion relations for the possible trajectory structures that allow us to prove two things.  The first is that, if we close the system, the semiclassical approximation preserves normalization as all higher order contributions cancel and the survival probability remains constant at 1.  The second is a proof that the full continuity equation is satisfied to all orders.  We also show that we can move via another continuity equation from the current density to a transport picture in terms of trajectories that start and end in the leads.  All of these results are valid for time scales up to the Heisenberg time, and the direct extension to times longer than the Heisenberg time remains an open problem.  We also restrict ourselves to a regime where the Ehrenfest time is much shorter than the average dwell time so that Ehrenfest time effects can be ignored.

This article is organised as follows.  We first introduce the semiclassical current density in section~\ref{semicurrdens}, where we also calculate the diagonal approximation, while in section~\ref{offdiag} we extend the semiclassical calculation to higher orders.  To begin our proof that the continuity equation holds to all orders, we first shift to the energy domain via a Fourier transform in section~\ref{fourier} and recall the calculation of the survival probability.  Our recursion relation arguments are presented in section~\ref{recursion}, along with the proof that both unitarity and the continuity equation are satisfied in the semiclassical approximation.  In section~\ref{transport} we explore the connection to quantum transport and end with our conclusions in section~\ref{conclusions}.

\section{The semiclassical current density} \label{semicurrdens}

In order to calculate a semiclassical approximation to the current density, we start by writing the wavefunction
\beq \label{psiteqn}
\psi(\r,t)=
\int \rmd \r' \: K(\r,\r',t)\psi_{\o}(\r'), 
\eeq
in terms of the quantum propagator $K({\bf r},{\bf r'},t)$ and the initial wavefunction $\psi_{\o}(\r')$ at time $t=0$.  We then replace the exact quantum propagator with the semiclassical Van Vleck propagator \cite{gutzwiller90}
\beq \label{smcpropeqn}
K^{\rm sc}\left(\r,\r',t\right)=\frac{1}{(2\pi\rmi\hbar)^{f/2}}\sum_{\tilde\gamma \left(\r'\to\r,t\right)}D_{\tilde\gamma }\rme^{\frac{\rmi}{\hbar}S_{\tilde\gamma}(\r,\r',t)},
\eeq
where $f$ is the dimension of the system, though in the following we will consider $f=2$.  $S_{\tilde \gamma}(\r,\r',t)$ is the action along the path $\tilde\gamma$ connecting $\r'$ and $\r$ in time $t$, and $D_{\tilde\gamma} =\left| \det -\partial^2 S_{\tilde \gamma}(\r,\r',t)/\partial \r \partial \r'\right|^{1/2}\exp\left(-\rmi\pi\mu_{\tilde\gamma}/2\right)$ is the Van Vleck determinant including the phase due to the Morse index $\mu_{\tilde\gamma}$.

Substituting \eref{psiteqn} and \eref{smcpropeqn} into the expression for the current density \eref{jrteqn}, we obtain
\beqa \label{jtrajeqn1}
\j^{\rm sc}(\r,t)=&\frac{1}{8m\pi^2 \hbar^2}\int_A \rmd\r'\rmd\r''\:\psi_{\o}(\r')\psi_{\o}^*(\r'')\\ \nonumber
&\times \sum_{\tilde\gamma \left(\r'\to\r,t\right)\atop \tilde\gamma' \left(\r''\to\r,t\right) } D_{\tilde\gamma }D_{\tilde\gamma' }^*\rme^{\frac{\rmi}{\hbar}(S_{\tilde\gamma} -S_{\tilde\gamma'})}\left[\p_{\tilde\gamma,\f}+{\p}_{\tilde\gamma',\f}\right],
\eeqa
which involves a sum over pairs of trajectories $\tilde\gamma$ and $\tilde\gamma'$, which have final momenta $\p_{\tilde\gamma,\f}$ and $\p_{\tilde\gamma',\f}$ at $\r$. Due to the highly oscillating phase most of the contributions will cancel out upon averaging (for example over a local time average, though we do not include this explicitly in our notation).  The remaining systematic contributions come from pairs of trajectories that are highly correlated and `similar'. As in \cite{waltner08, gutierrez08}, we neglect changes in the slowly varying prefactor and we expand the trajectories $\tilde\gamma$ and $\tilde\gamma'$, which go from $\mathbf{r}'$ and $\mathbf{r}''$ respectively to $\mathbf{r}$ in time $t$, around trajectories $\gamma$ and $\gamma'$ which travel from the midpoint $\mathbf{r}_\o=(\mathbf{r}'+\mathbf{r}'')/2$ to $\mathbf{r}$ also in a time $t$.  We can then express \eref{jtrajeqn1} as
\beq \label{jtrajeqn2}
\fl \j^{\rm sc}(\r,t)=\frac{1}{4m\pi^2 \hbar^2}\int \rmd \r_{\o}\sum_{\gamma,\gamma'(\r_{\o}\to \r,t)}D_{\gamma}D_{\gamma'}^*\rme^{\frac{\rmi}{\hbar}(S_{\gamma}-S_{\gamma'})}\rho_{\mathrm{W}}(\r_{\o},\p_{\gamma\gamma'}^{\o})\p_{\gamma\gamma'}^{\f},
\eeq
where $\p_{\gamma\gamma'}^{\o}$ is the average initial momentum and $\p_{\gamma\gamma'}^{\f}$ the average final momentum of the trajectories $\gamma$ and $\gamma'$, and 
\beq
\rho_{\mathrm{W}}(\r_{\o},\p)=\int\rmd \q \: \psi_{\o}\left(\r_{\o}+\frac{\q}{2}\right)\psi_{\o}^*\left(\r_{\o}-\frac{\q}{2}\right)\rme^{-\frac{\rmi}{\hbar} \q\cdot\p},
\eeq
is the Wigner transform of $\psi_{\o}(\r_{\o})$, which arises from setting $\q=(\r'-\r'')$.  This current density, or more accurately the integrated current density
\beq
J(t)=\int _{S} \j(\r,t)\cdot \hat{n}_x \rmd x,
\eeq
is the quantity we wish to evaluate semiclassically.  The integral over the cross-section of the lead means that we are interested in trajectory pairs that start inside the system and end in the lead itself.

\subsection{Diagonal approximation} \label{diagonal}

The simplest semiclassical contribution to calculate is the diagonal approximation \cite{berry85} where we pair the trajectories with themselves, $\gamma=\gamma'$ .  Restricting ourselves to these pairs, \eref{jtrajeqn2} simplifies to
\beq \label{jtrajdiageqn1}
\j^{\rm diag}(\r,t)\cdot \hat{n}_x = \frac{1}{4m\pi^2 \hbar^2}\int \rmd \r_{\o}\sum_{\gamma(\r_{\o}\to\r,t)}|D_{\gamma}|^2 \rho_{\mathrm{W}}(\r_{\o},\p_{\gamma,\o})p_{x,\gamma}^{\f},
\eeq
while the same treatment for the probability density leads to the similar result of
\beq \label{rhortrajdiageqn1}
\rho^{\rm diag}(\r,t)= \frac{1}{4\pi^2 \hbar^2}\int \rmd \r_{\o}\sum_{\gamma(\r_{\o}\to\r,t)}|D_{\gamma}|^2 \rho_{\mathrm{W}}(\r_{\o},\p_{\gamma,\o}).
\eeq

Performing the sum over trajectories in \eref{jtrajdiageqn1} using the open sum rule \cite{rs02,ha84,sieber99}, the diagonal approximation becomes
\beq \label{jtrajdiageqn2}
 \j^{\rm diag}(\r,t)\cdot \hat{n}_x= \left\langle\frac{\mu}{w}\rme^{-\mu t}\right\rangle_{\r,\p}, \qquad \r \in S,
\eeq
where $w$ is the size of the opening, $\mu=1/\tau_d$ is the classical escape where $\tau_d$ is the dwell time, and $\langle\dots\rangle_{\r,\p}$ is a phase space average.

Integrating over the opening cross-section just leads to a factor $w$ so that
\beq \label{Jdiageqn}
J^{\rm diag}(t) = \mu \rme^{-\mu t},
\eeq
where by supposing that the wave function has a well defined energy we can drop the average over phase space.  By integrating with respect to time and setting $\rho^{\rm diag}(0)=1$, we obtain 
\beq \label{rhodiageqn}
\rho^{\rm diag}(t)=e^{-\mu t},
\eeq
which is the classical decay for a chaotic system for long times.  This result also follows directly from \eref{rhortrajdiageqn1} by using the open sum rule and integrating over the volume of the system \cite{waltner08, gutierrez08}.

\section{Off-diagonal contributions} \label{offdiag}

\begin{figure}
\centering
$\begin{array}{ll}
(a) & (b) \\
\hspace{0.15cm} \includegraphics[width=6cm]{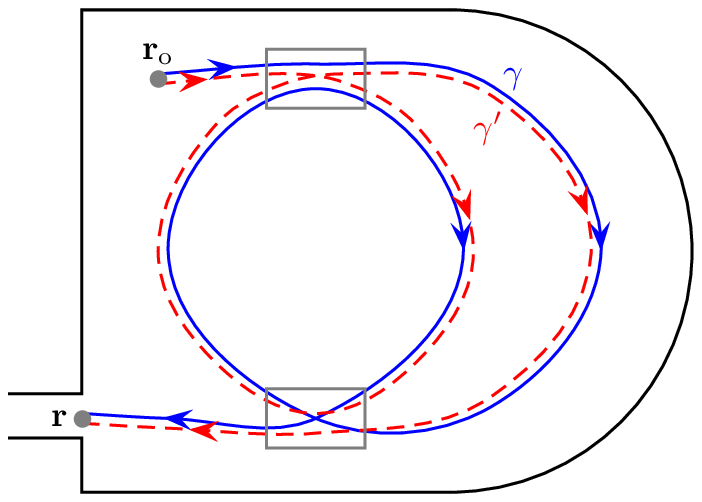} & \hspace{0.15cm} \includegraphics[width=6cm]{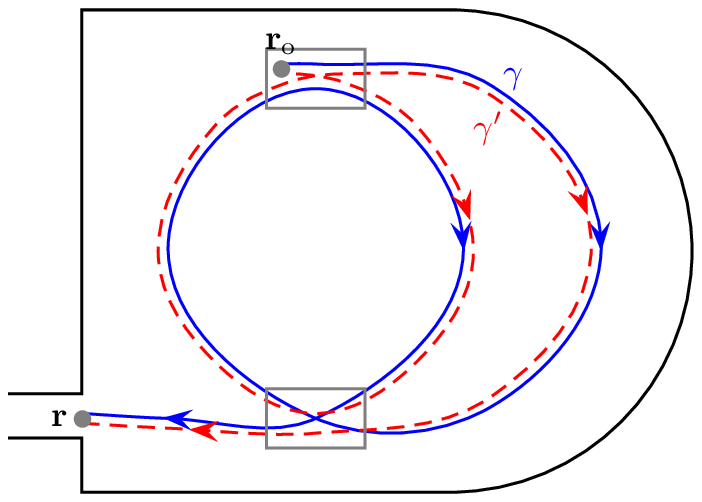}
\end{array}$
\caption{A schematic picture of a trajectory $\gamma$ (solid line) that approaches itself twice in two self-encounters, and a partner trajectory $\gamma'$ (dashed line) formed by crossing the encounters differently.  In (a) the start of the trajectories is outside of the first encounter (2ll) while in (b) it has been shifted inside the encounter (1ll).}
\label{trajpic}
\end{figure}

To calculate higher order corrections we consider the contributions of trajectories that have close self-encounters.  Highly correlated partner trajectories can then be found that follow the original trajectory almost exactly, but which differ in the encounter regions leading to a small action difference. An example of such a trajectory pair is given in Figure~\ref{trajpic}a.

We are interested in calculating the contributions of trajectories with an arbitrary number of encounters, each of an arbitrary size, and we work along the same lines as \cite{muller04,muller05,heusler06,muller07}.  An encounter that involves $l$ stretches of the trajectory is called an $l$-encounter, so that the example in Figure~\ref{trajpic}a has two 2-encounters.  Trajectories can be labelled by the vector $\v$, whose elements $v_l$ list the number of $l$-encounters along the trajectory (and hence also along its partner).  The total number of encounters is $V=\sum v_{l}$, while inbetween the encounters are long trajectory stretches called links.  We make use of the fact that we can generate the possible configurations (or structures) of trajectories labelled by $\v$ from the related closed orbit structures (formed by connecting the start and the end of the trajectory).  The number of links of the related closed orbit is $L=\sum lv_{l}$, and by cutting each of the links in turn (and moving the cut ends to the required positions) we generate the open trajectories.  The number of trajectory structures $N(\v)$ with the same $\v$ is simply related to the number of closed periodic orbit structures (in fact for the closed orbits, we effectively count the same orbit $L$ times and have to divide by that factor).  When we cut a link of the closed periodic orbit, this generates two links in the trajectory structure, so that the total number of links is $L+1$.  However, this usual contribution \cite{heusler06,muller07}, which are also called `two-leg-loops' (2ll) in \cite{waltner08}, is just one of the possible contributions.  We can also shrink the link at the start of the trajectory so that the start moves into the first encounter, as in Figure~\ref{trajpic}b.  This case gives a different semiclassical contribution and corresponds to the `one-leg-loops' (1ll) of \cite{waltner08}.  We note that because the trajectory leaves the system at its end (which is placed in the lead) it cannot then return to a nearby point to have an encounter, and so we can only shrink the link at the start of the trajectory.  Therefore the possibility that both the start and end point are inside encounters, the 0ll case in \cite{gutierrez08}, cannot happen. As such, the semiclassical contribution can be separated into these two cases:
\begin{description}
\item[A] where the start point is outside of the encounters (2ll),
\item[B] where the start point is inside an encounter (1ll) 
\end{description}

\subsection{Case A: two-leg-loops}

For structures corresponding to a vector $\v$, the action difference between the trajectories $\gamma$ and $\gamma'$ is given by $\Delta S=\s\u$ in the linearized approximation.  The vectors $\s$ and $\u$ contain the differences, along the stable and unstable manifold respectively, of the encounter stretches (of $\gamma$) in Poincar\'e surfaces transverse to each encounter.  Their semiclassical contribution can be evaluated using an auxiliary weight, $w_{\v,\mathrm{A}}(\s,\u,t)$, of such encounters
\beq \label{JAcontribeqn}
J_{\v, \mathrm{A}}(t)=\mu N(\v)\int \rmd\s\:\rmd\u\:w_{\v,\mathrm{A}}(\s,\u,t)\rme^{-\mu t_{\mathrm{exp}}}\rme^{\frac{\rmi}{\hbar}\s\u},
\eeq
where $N(\v)$ is the number of trajectory structures corresponding to the vector $\v$.  The exponential $\rme^{-\mu t_{\mathrm{exp}}}$ is the average survival probability of the structures and involves a correction due to the proximity of encounter stretches during the encounters.  Because of this proximity, if one encounter stretch survives then they all should do and so the exposure time $t_{\mathrm{exp}}$ involves only the time of a single traversal of each encounter.  If we label the $V$ encounters by $\alpha$, with each involving $l_{\alpha}$ encounter stretches that last $t_{\mathrm{enc}}^{\alpha}$, and we label the $L+1$ links by $i$, with each lasting $t_{i}$, then the exposure time is simply
\beq \label{exptimeeqn}
t_{\mathrm{exp}}=\sum_{i=1}^{L+1}t_{i}+\sum_{\alpha=1}^{V}t_{\mathrm{enc}}^{\alpha}= t - \sum_{\alpha=1}^{V}(l_{\alpha}-1)t_{\mathrm{enc}}^{\alpha}.
\eeq
Each encounter time is defined as the the time during which the stable and unstable separations of all the encounter stretches remain smaller than some classical constant $c$.  They are therefore given by
\beq
t_{\mathrm{enc}}^{\alpha}=\frac{1}{\lambda}\ln \frac{c^2}{\mathrm{max}_j\vert s_{\alpha,j}\vert \times \mathrm{max}_j\vert u_{\alpha,j}\vert},
\eeq
where $\lambda$ is the Lyapunov exponent, while $\s_{\alpha}$ and $\u_{\alpha}$ are vectors containing only the stable and unstable separations of the encounter $\alpha$.

The weight of encounters can be expressed in terms of an integral
\beq \label{weightinttrajeqnA}
w_{\v,\mathrm{A}}(\s,\u,t)=\frac{\int_{0}^{t-t_{\mathrm{enc}}}\rmd t_{L}\ldots\int_{0}^{t-t_{\mathrm{enc}}-t_{L}\ldots-t_{2}}\rmd t_{1}}{\Omega^{L-V}\prod_{\alpha}t_{\mathrm{enc}}^{\alpha}},
\eeq
where $t_{\mathrm{enc}}$ is the total time the trajectory spends in the encounters, $t_{\mathrm{enc}}=\sum_{\alpha=1}^{V}l_{\alpha}t_{\mathrm{enc}}^{\alpha}$.  The links should all have positive duration, and the weight is simply an $L$-fold integral over the first $L$ link times, while the last link time is fixed by the total trajectory time $t$.  When we perform the integrals the weight function becomes
\beq \label{weighttrajeqnA}
w_{\v,\mathrm{A}}(\s,\u,t)=\frac{\left(t-\sum_{\alpha}l_{\alpha}t_{\mathrm{enc}}^{\alpha}\right)^L}{L!\Omega^{L-V}\prod_{\alpha}t_{\mathrm{enc}}^{\alpha}}.
\eeq

To calculate the semiclassical contribution we will rewrite \eref{JAcontribeqn} as
\beq
J_{\v, \mathrm{A}}(t)=\mu N(\v)\int \rmd\s\:\rmd\u\:z_{\v,\mathrm{A}}(\s,\u,t)\rme^{-\mu t}\rme^{\frac{\rmi}{\hbar}\s\u},
\eeq
where $z_{\v,\mathrm{A}}(\s,\u,t)$ is an augmented weight including the term from the survival probability correction of the encounters
\beqa \label{zeqnA}
z_{\v,\mathrm{A}}(\s,\u,t)&=w_{\v,\mathrm{A}}(\s,\u,t)\rme^{\sum_{\alpha}(l_{\alpha}-1)\mu t_{\mathrm{enc}}^{\alpha}} \\ \nonumber
&\approx\frac{\left(t-\sum_{\alpha}l_{\alpha}t_{\mathrm{enc}}^{\alpha}\right)^L\prod_{\alpha}\left(1+(l_{\alpha}-1)\mu t_{\mathrm{enc}}^{\alpha}\right)}{L!\Omega^{L-V}\prod_{\alpha}t_{\mathrm{enc}}^{\alpha}},
\eeqa
where we have expanded the exponent to first order.  The semiclassical contribution comes from terms where the encounter times in the numerator cancel those in the denominator exactly \cite{muller04,muller05}.  Keeping only those terms, we then obtain a factor of $(2\pi\hbar)^{L-V}$ from the integrals over $\s$ and $\u$.  Finally we need the number of structures $N(\v)$ corresponding to each vector, which are tabulated in \cite{muller05b}, and we can obtain the result for trajectories described by the vector $\v$ of interest.

As an example, in Table~\ref{trajtableA}, we calculate the contribution of trajectory structures described by vectors $\v$ with $L-V=3$, for systems with time reversal symmetry.  We use a shorthand notation to describe the encounters in the vector $\v$ so that each term $(l)^{v_{l}}$ represents that the structure has $v_l$ $l$-encounters. 
\Table{\label{trajtableA}Contribution A of different types of trajectory pairs to the integrated current density, along with the number of structures for systems with time reversal symmetry.}
\begin{tabular}{ccccc}
\br
$\v$&$L$&$V$&$\frac{J_{\v, \mathrm{A}}(t)}{N(\v)}$&$N(\v)$\\
\mr
$(2)^{3}$&6&3&$\frac{\mu\rme^{-\mu t}}{T_{\mathrm{H}}^3}\left(-\frac{4t^{3}}{3}+\frac{\mu t^{4}}{2}-\frac{\mu^2 t^{5}}{20}+\frac{\mu^3 t^{6}}{720}\right)$&41\\
$(2)^{1}(3)^1$&5&2&$\frac{\mu\rme^{-\mu t}}{T_{\mathrm{H}}^3}\left(t^{3}-\frac{7\mu t^{4}}{24}+\frac{\mu^2 t^{5}}{60}\right)$&60\\
$(4)^{1}$&4&1&$\frac{\mu\rme^{-\mu t}}{T_{\mathrm{H}}^3}\left(-\frac{2t^{3}}{3}+\frac{\mu t^{4}}{8}\right)$&20\\
\br
\end{tabular}
\endTable
These results can be multiplied by the number of structures and summed to give the third order correction for this case
\beq
J_{3, \mathrm{A}}(t) = \frac{\mu\rme^{-\mu t}}{T_{\mathrm{H}}^3}\left(-8t^{3}+\frac{11\mu t^{4}}{2}-\frac{21\mu^2 t^{5}}{20}+\frac{41\mu^3 t^{6}}{720}\right).
\eeq

\subsection{Case B: one-leg-loops}

We can write this contribution as
\beq
J_{\v, \mathrm{B}}(t)=\mu N(\v)\int \rmd\s\:\rmd\u\:z_{\v,\mathrm{B}}(\s,\u,t)\rme^{-\mu t}\rme^{\frac{\rmi}{\hbar}\s\u}.
\eeq
Now that one encounter overlaps with the start of the trajectory, we have one link fewer ($L$ in total) and an extra integral over the position of the encounter relative to the starting point.  Starting with a closed periodic orbit (and dividing by the overcounting factor of $L$), we can cut each of the $L$ links in turn and move the encounter on one side of the cut to the start.  In total we obtain $l_{\alpha'}$ copies of the same 1ll involving the encounter $\alpha'$.  The augmented weight can then be expressed as a sum over the different possibilities, each of which involves an integral over the position of the start point inside the encounter, $t_{\alpha'}$, \cite{waltner08}
\beq \label{ztrajeqnB}
\fl z_{\v,\mathrm{B}}(\s,\u,t)=\sum_{\alpha'=1}^{V}l_{\alpha'}\int_{0}^{\frac{1}{\lambda}\ln\frac{c}{\mathrm{max}_j\vert s_{\alpha',j}\vert}} \rmd t_{\alpha'}\frac{\left(t-\sum_{\alpha}l_{\alpha}t_{\mathrm{enc}}^{\alpha}\right)^{L-1}}{L!\Omega^{L-V}\prod_{\alpha}t_{\mathrm{enc}}^{\alpha}}\rme^{\sum_{\alpha}(l_{\alpha}-1)\mu t_{\mathrm{enc}}^{\alpha}},
\eeq
where the time of encounter $\alpha'$ is related to the starting position via 
\beq
t_{\mathrm{enc}}^{\alpha'}=t_{\alpha'}+\frac{1}{\lambda}\ln\frac{c}{\mathrm{max}_j\vert u_{\alpha',j}\vert}.
\eeq
Because of the integrals over the position of the start point inside the encounter, the semiclassical contribution is calculated differently, using integrals of the type in \cite{br06}.  With a change of variables however \cite{gutierrez08}, the integral effectively gives a factor of $t_{\mathrm{enc}}^{\alpha'}$, and we can write the augmented weight as
\beq \label{zeqnB}
\fl z_{\v,\mathrm{B}}(\s,\u,t)\approx\frac{\left(\sum_{\alpha}l_{\alpha}t_{\mathrm{enc}}^{\alpha}\right)\left(t-\sum_{\alpha}l_{\alpha}t_{\mathrm{enc}}^{\alpha}\right)^{L-1}\prod_{\alpha}\left(1+(l_{\alpha}-1)\mu t_{\mathrm{enc}}^{\alpha}\right)}{L!\Omega^{L-V}\prod_{\alpha}t_{\mathrm{enc}}^{\alpha}},
\eeq
and treat it as before.

Continuing with our example, in Table~\ref{trajtableB}, we calculate this type of contribution for trajectories with $L-V=3$, for systems with time reversal symmetry.
\Table{\label{trajtableB}Contribution B of different types of trajectory pairs to the current density for systems with time reversal symmetry.}
\begin{tabular}{ccccc}
\br
$\v$&$L$&$V$&$\frac{J_{\v, \mathrm{B}}(t)}{N(\v)}$&$N(\v)$\\
\mr
$(2)^{3}$&6&3&$\frac{\mu\rme^{-\mu t}}{T_{\mathrm{H}}^3}\left(\frac{2t^{3}}{3}-\frac{\mu t^{4}}{6}+\frac{\mu^2 t^{5}}{120}\right)$&41\\
$(2)^{1}(3)^1$&5&2&$\frac{\mu\rme^{-\mu t}}{T_{\mathrm{H}}^3}\left(-\frac{2t^{3}}{5}+\frac{7\mu t^{4}}{120}\right)$&60\\
$(4)^{1}$&4&1&$\frac{\mu\rme^{-\mu t}}{T_{\mathrm{H}}^3}\left(\frac{t^{3}}{6}\right)$&20\\
\br
\end{tabular}
\endTable
These results can be multiplied by the number of structures and summed to give the third order correction for this case
\beq
J_{3, \mathrm{B}}(t) = \frac{\mu\rme^{-\mu t}}{T_{\mathrm{H}}^3}\left(\frac{20t^{3}}{3}-\frac{10\mu t^{4}}{3}+\frac{41\mu^2 t^{5}}{120}\right).\eeq

\subsection{Results}

By simply adding the results for case A and case B, we can obtain the results for each vector and for each symmetry class.  However, as can be seen in the example above, the Heisenberg time dependence involves only the value of $L-V$ of the vector, and we can further sum over all vectors $\v$ with the same value of $L-V$ to obtain that order in the expansion of $J(t)$.  The perturbative expansion is therefore in powers of $t/T_{\mathrm{H}}$, where each term involves a finite expansion in powers of $\mu t$.  We note that the same ordering of structures for transport quantities like the conductance leads to an expansion in powers of the inverse number of channels \cite{heusler06}.  When we perform the expansion of the current density for the unitary case we obtain
\numparts
\beqa
J_{2}(t) &= \frac{\mu\rme^{-\mu t}}{T_{\mathrm{H}}^2}\left(-\frac{\mu t^{3}}{6}+\frac{\mu^2 t^{4}}{24}\right),\\
J_{4}(t) &= \frac{\mu\rme^{-\mu t}}{T_{\mathrm{H}}^4}\left(-\frac{\mu t^{5}}{15}+\frac{\mu^2 t^{6}}{20}-\frac{7\mu^3 t^{7}}{720}+\frac{\mu^4 t^{8}}{1920}\right),\\
J_{6}(t) &= \frac{\mu\rme^{-\mu t}}{T_{\mathrm{H}}^6}\left(-\frac{\mu t^{7}}{28}+\frac{401\mu^2 t^{8}}{10080}-\frac{643\mu^3 t^{9}}{45360}+\frac{2\mu^4 t^{10}}{945}\right. \\  \nonumber
& \qquad\qquad\left. -\frac{11\mu^5 t^{11}}{80640}+\frac{\mu^6 t^{12}}{322560}\right),
\eeqa
\endnumparts
and zero for odd values of $L-V$.  For the orthogonal case we obtain
\numparts
\beqa
J_{1}(t) &= \frac{\mu\rme^{-\mu t}}{T_{\mathrm{H}}}\left(-t+\frac{\mu t^{2}}{2}\right),\\
J_{2}(t) &= \frac{\mu\rme^{-\mu t}}{T_{\mathrm{H}}^2}\left(t^{2}-\frac{7\mu t^{3}}{6}+\frac{5\mu^2 t^{4}}{24}\right),\\
J_{3}(t) &= \frac{\mu\rme^{-\mu t}}{T_{\mathrm{H}}^3}\left(-\frac{4t^{3}}{3}+\frac{13\mu t^{4}}{6}-\frac{17\mu^2 t^{5}}{24}+\frac{41\mu^3 t^{6}}{720}\right),\\
J_{4}(t) &= \frac{\mu\rme^{-\mu t}}{T_{\mathrm{H}}^4}\left(2t^{4}-\frac{39\mu t^{5}}{10}+\frac{43\mu^2 t^{6}}{24}-\frac{197\mu^3 t^{7}}{720}+\frac{509\mu^4 t^{8}}{40320}\right),\\
J_{5}(t) &= \frac{\mu\rme^{-\mu t}}{T_{\mathrm{H}}^5}\left(-\frac{16t^{5}}{5}+\frac{106\mu t^{6}}{15}-\frac{121\mu^2 t^{7}}{30}+\frac{61\mu^3 t^{8}}{70}\right. \\  \nonumber 
& \qquad\qquad\left. -\frac{1321\mu^4 t^{9}}{17280}+\frac{2743\mu^5 t^{10}}{1209600}\right),\\
J_{6}(t) &= \frac{\mu\rme^{-\mu t}}{T_{\mathrm{H}}^6}\left(\frac{16t^{6}}{3}-\frac{5419\mu t^{7}}{420}+\frac{86801\mu^2 t^{8}}{10080}-\frac{53273\mu^3 t^{9}}{22680}\right. \\ \nonumber 
& \qquad\qquad\left. +\frac{1699\mu^4 t^{10}}{5760}-\frac{4063\mu^5 t^{11}}{241920}+\frac{55459\mu^6 t^{12}}{159667200}\right).
\eeqa
\endnumparts

By direct comparison with the results obtained for the survival probability in \cite{gutierrez08} we can see that, term by term, 
\beq
\frac{\partial}{\partial t}\rho_{m}(t)=-J_{m}(t),
\eeq
for both symmetry classes and all $m\leq6$.  Up to this order, it then follows trivially by summing over $m$ that the continuity equation \eref{intconteqn} is satisfied.  We will prove that this result holds to all orders, and in order to do so we first consider the (inverse) Fourier transform of both the current density and the survival probability.

\section{Fourier transform}\label{fourier}

To show that the continuity equation holds to all orders it is simpler to make a Fourier transform rather than showing, for each $m$, that all terms sum to zero for each power of $t$.  In fact this is reminiscent of the situation for parametric correlations for systems without time reversal symmetry where it is simpler to show agreement with RMT (for times shorter than the Heisenberg time) to all orders for the correlation function \cite{nagao07} rather than for the form factor \cite{ks07a}.

For convenience we restrict ourselves to positive times, and we will consider the (one-sided) inverse Fourier transform of the current density $J(t)$
\beq
\J(\omega)=\int_{0}^{\infty} \rmd \tau \: J(\tau T_{\mathrm{H}}) \rme^{2\pi\rmi\omega\tau},
\eeq
as well as of the survival probability $\rho(t)$
\beq
P(\omega)=\int_{0}^{\infty} \rmd \tau \: \rho(\tau T_{\mathrm{H}}) \rme^{2\pi\rmi\omega\tau},
\eeq
where $\tau=t/T_{\mathrm{H}}$, and we will also use the number of open channels $M=\mu{T_{\mathrm{H}}}$.  In the Fourier space the continuity equation \eref{intconteqn} becomes
\beq \label{fourierconteq}
T_{\mathrm{H}}\J(\omega)-(2\pi\rmi\omega)P(\omega)=1,
\eeq
and this is the relation which we want to show semiclassically.  We note that the 1 on the right hand side comes from the diagonal terms given in \eref{Jdiageqn} and \eref{rhodiageqn}. 

\subsection{Transformed current density}

Interestingly the semiclassical contribution for the current density can be separated into a product of contributions over the encounters and the links of the trajectory, in a similar way as for the conductance \cite{heusler06}.  This means that we can obtain the semiclassical result very simply.  For example for case A the contribution from trajectories with structures described by $\v$ is
\beqa
\J_{\v,\mathrm{A}}(\omega)=&\frac{\mu N(\v)}{T_{\mathrm{H}}} \left(\prod_{i=1}^{L+1}\int_{0}^{\infty} \rmd t_{i} \: \rme^{-\left(\mu-\frac{2\pi\rmi\omega}{T_{\mathrm{H}}}\right) t_{i}}\right) \\ \nonumber 
& \times \left(\prod_{\alpha=1}^{V}\int \rmd\s_{\alpha}\rmd\u_{\alpha}\:\frac{\rme^{-\left(\mu-\frac{2\pi\rmi\omega l_{\alpha}}{T_{\mathrm{H}}}\right) t_{\mathrm{enc}}^{\alpha}}\rme^{\frac{\rmi}{\hbar}\s_{\alpha}\u_{\alpha}}}{\Omega^{l_{\alpha}-1}t_{\mathrm{enc}}^{\alpha}}\right),
\eeqa
where we have used the first expression for the exposure time in \eref{exptimeeqn}.  We can perform these integrals as before, and because the Heisenberg times mutually cancel, we effectively get a factor of $(M-2\pi\rmi\omega)^{-1}$ for each link and a factor of $-(M-2\pi\rmi\omega l_{\alpha})$ for each encounter.  The contribution therefore simplifies to
\beq \label{JvAeqn}
\J_{\v,\mathrm{A}}(\omega)=\mu N(\v)(-1)^V \frac{\prod_{\alpha}(M-2\pi\rmi\omega l_{\alpha})}{(M-2\pi\rmi\omega)^{L+1}}.
\eeq

Considering case B, we have one link fewer ($L$ in total) and one encounter, $\alpha'$ at the start of the trajectory pair.  The integral over $\s_{\alpha'}$ and $\u_{\alpha'}$ is of course different from the integrals over the remaining encounters but we effectively obtain a factor of $1$.  We also remember that to obtain the structures for this case we started from the corresponding closed orbits, so that we divide by the overcounting factor $L$, and note that the encounter $\alpha'$ occurs at the start $l_{\alpha'}$ times.  Altogether, the contribution simplifies to
\beq
\J_{\v,\mathrm{B}}(\omega)=\frac{\mu N(\v)(-1)^{V-1}}{L}\sum_{\alpha'}l_{\alpha'} \frac{\prod_{\alpha\neq\alpha'}(M-2\pi\rmi\omega l_{\alpha})}{(M-2\pi\rmi\omega)^{L}},
\eeq
which we can also write as
\beq \label{JvBeqn}
\fl \J_{\v,\mathrm{B}}(\omega)=\frac{\mu N(\v)(-1)^{V-1}}{L}\left(\sum_{\alpha'}\frac{l_{\alpha'}}{(M-2\pi\rmi\omega l_{\alpha'})}\right)\frac{\prod_{\alpha}(M-2\pi\rmi\omega l_{\alpha})}{(M-2\pi\rmi\omega)^{L}},
\eeq
to remove the restriction on the product over $\alpha$.

\subsection{Transformed survival probability}

For the survival probability, as we saw in \cite{gutierrez08}, because the trajectories that contribute start and end inside the cavity we additionally have the possibility that there is an encounter that overlaps with the end of the trajectory leading to three possible cases:

\begin{description}
\item[A] where the start and end points are outside of the encounters (2ll),
\item[B] where either the start or the end point is inside an encounter (1ll) and 
\item[C] where both the start and end point are inside encounters (0ll).
\end{description}

We can again write the semiclassical contribution as a product of contributions from the links and encounters and for the first two cases we obtain very similar results as for the integrated current density in equations \eref{JvAeqn} and \eref{JvBeqn}.  For case A we arrive at
\beq \label{PvAeqn}
P_{\v,\mathrm{A}}(\omega)=\frac{\J_{\v,\mathrm{A}}(\omega)}{\mu}.
\eeq
For case B, because the end of the trajectory is inside the system it can also be during an encounter.  We therefore obtain an additional factor of 2 compared to the integrated current density and the result is then
\beq \label{PvBeqn}
P_{\v,\mathrm{B}}(\omega)=2\frac{\J_{\v,\mathrm{B}}(\omega)}{\mu}.
\eeq

For case C things are slightly more complicated as we now have both ends of our trajectories inside different encounters and again one link fewer.  The two encounters at the ends effectively give factors of $1$ while the remaining encounters and links give their usual contributions.  We also need to know the number of possible structures.  Starting again with the corresponding closed periodic orbit structures described by the same vector $\v$ (and dividing by the overcounting factor $L$) we can generate the required trajectories by cutting the links which connect two different encounters $\alpha'$ and $\beta'$.  We can then shrink the cut ends and shift them to the correct positions to obtain the trajectory structures.  If we record in the matrix elements $\N_{\alpha,\beta}(\v)$ the number of ways of cutting links that connect encounters $\alpha$ and $\beta$ in all the periodic orbit structures described by $\v$ then we can write the contribution as
\beq \label{PvCeqn}
\fl P_{\v,\mathrm{C}}(\omega)=\left(\sum_{\alpha,\beta}\frac{\N_{\alpha,\beta}(\v)}{(M-2\pi\rmi\omega l_{\alpha})(M-2\pi\rmi\omega l_{\beta})}\right)\frac{(-1)^{V-2}\prod_{\alpha}(M-2\pi\rmi\omega l_{\alpha})}{(M-2\pi\rmi\omega)^{L-1}}.
\eeq

\section{Recursion relations} \label{recursion}

In the limit $\mu\to0$ (or $\tau_d\to\infty$) the quantum wavepacket remains inside the cavity and the survival probability is identically 1 for all times.  This term comes from the diagonal approximation, which means that all higher order terms from correlated trajectories should vanish.  As was shown for the survival probability, $\rho(t)$, in \cite{waltner08,gutierrez08} one-leg-loops (cases B and C) are necessary to ensure this unitarity and cancel the terms from two-leg-loops (case A) which do not vanish when the system is closed.  We first show that this holds to all orders by developing recursion relations between the different types of contribution.

\subsection{Unitarity}\label{recursionunitarity}

By considering the sum over vectors $\v$ which share the same value of $L-V=m$, and summing over the different cases, we will show that $P_{m}(\omega,\mu=0)$ vanishes for all $m>0$ and both symmetry classes.  This is equivalent to showing that the lowest order $t$ term in the polynomial multiplying $\exp(-\mu t)$ of the various $\rho_{\v}(t)$ also sum to zero, so that $\rho_{m}(t)$ involves a polynomial with lowest term proportional to $t^{m+1}$.  We start with the semiclassical results for each case when $\mu=0$, and for case A we have
\beq
P_{\v,\mathrm{A}}(\omega,\mu=0)= N(\v)\frac{(-1)^{L+1}\prod_{\alpha}l_{\alpha}}{(2\pi\rmi\omega)^{L-V+1}},
\eeq
while similarly for case B, the result is
\beq
P_{\v,\mathrm{B}}(\omega,\mu=0)= N(\v)\frac{2V}{L} \frac{(-1)^{L}\prod_{\alpha}l_{\alpha}}{(2\pi\rmi\omega)^{L-V+1}},
\eeq
because $\sum_\alpha 1 = V$.  For case C we obtain
\beq
P_{\v,\mathrm{C}}(\omega,\mu=0)=\left(\sum_{\alpha,\beta}\frac{\N_{\alpha,\beta}(\v)}{l_{\alpha}l_{\beta}}\right)\frac{(-1)^{L-1}\prod_{\alpha} l_{\alpha}}{(2\pi\rmi\omega)^{L-V+1}}.
\eeq
For this case, it is useful to rewrite the sum over $\alpha$ and $\beta$ as a sum over the components of the vector $\v$.  $\N_{\alpha,\beta}(\v)$ records the number of ways of cutting links, that connect encounters $\alpha$ and $\beta$, in the periodic orbits described by $\v$.  However we can see that the important quantities are the sizes of encounters $\alpha$ and $\beta$.  Instead we record in $\N_{k,l}(\v)$ the number of links that join an encounter of size $k$ to an encounter of size $l$.

With this in place we can sum the different types of contributions, and then sum over vectors $\v$ with the same $L-V=m$
\beqa \label{Pmmuzeroeqn}
\fl P_{m}(\omega,\mu=0)=\frac{(-1)^{m+1}}{(2\pi\rmi\omega)^{m+1}}\sum_{\v}^{L-V=m}(-1)^{V}\prod_{\alpha}l_{\alpha}\left[\left(1-\frac{2V}{L}\right)N(\v)+\sum_{k,l}\frac{\N_{k,l}(\v)}{kl}\right].&&\nonumber\\
&&
\eeqa
Let us first consider the third contribution
\beq \label{PmCmuzeroeqn}
P_{m,\mathrm{C}}(\omega,\mu=0)=\frac{(-1)^{m+1}}{(2\pi\rmi\omega)^{m+1}}\sum_{\v}^{L-V=m}(-1)^{V}\prod_{\alpha}l_{\alpha}\sum_{k,l}\frac{\N_{k,l}(\v)}{kl}.
\eeq
To simplify this further we use the following recursion relation which can be deduced from \cite{muller05, muller05b}.  We have that $\N_{k,l}(\v)$ records the number of links that join a $k$-encounter to an $l$-encounter.  Instead of cutting the link to make the trajectory structure, we imagine shrinking the link so that the $k$ and $l$-encounters merge to form a new $(k+l-1)$-encounter.  By considering the number of ways that it is possible to shrink the link and create a smaller periodic orbit structure, we obtain the result
\beq \label{twoencorbeqiveqn}
\N_{k,l}(\v)=\frac{(k+l-1)(v_{k+l-1}+1)}{L-1}N(\v^{[k,l\to k+l-1]}),
\eeq
where $v_{k+l-1}$ is the $(k+l-1)$-th component of $\v$ and $\v^{[k,l\to k+l-1]}$ is the vector obtained by decreasing the components $v_k$ and $v_l$ by one and increasing the component $v_{k+l-1}$ by one (so that $v_{k+l-1}+1=v_{k+l-1}^{[k,l\to k+l-1]}$).  We also use the substitution
\beq
\tilde{N}(\v)=\frac{(-1)^{V}\prod_{\alpha}l_{\alpha}}{L}N(\v),
\eeq
which allows us to rewrite \eref{twoencorbeqiveqn} as
\beq \label{twoencorbeqiveqn2}
(-1)^{V}\prod_{\alpha}l_{\alpha}\frac{\N_{k,l}(\v)}{kl}=-v_{k+l-1}^{[k,l\to k+l-1]}\tilde{N}(\v^{[k,l\to k+l-1]}),
\eeq
so that \eref{PmCmuzeroeqn} becomes
\beq \label{PmCmuzeroeqn2}
\fl P_{m,\mathrm{C}}(\omega,\mu=0)=-\frac{(-1)^{m+1}}{(2\pi\rmi\omega)^{m+1}}\sum_{\v}^{L-V=m}\sum_{k,l}v_{k+l-1}^{[k,l\to k+l-1]}\tilde{N}(\v^{[k,l\to k+l-1]}).
\eeq
As combining a $k$ and $l$-encounter reduces both $L$ and $V$ by one, the resulting vector $\v^{[k,l\to k+l-1]}$ still has the same value of $L-V=m$ and by considering $\v^{[k,l\to k+l-1]}$ as a dummy variable $\v'$ it can be shown \cite{muller05, muller05b} that 
\beq
\sum_{\v}^{L-V=m}v_{k+l-1}^{[k,l\to k+l-1]}\tilde{N}(\v^{[k,l\to k+l-1]})=\sum_{\v'}^{L-V=m}v'_{k+l-1}\tilde{N}(\v').
\eeq
By identifying this dummy vector $\v'$ with $\v$ and substituting into \eref{PmCmuzeroeqn2}, the total contribution to $P_{m}(\omega,\mu=0)$ in equation \eref{Pmmuzeroeqn} simplifies to
\beq \label{Pmmuzeroeqn2}
\fl P_{m}(\omega,\mu=0)=\frac{(-1)^{m+1}}{(2\pi\rmi\omega)^{m+1}}\sum_{\v}^{L-V=m}\left[(L-2V)-\sum_{k,l}v_{k+l-1}\right]\tilde{N}(\v).
\eeq
Concentrating on the sum over $k$ and $l$, we define $k'=k+l-1$ where $k'>l$ because $k\geq2$. The sum then becomes
\beqa \label{compsumeqn}
\sum_{k,l}v_{k+l-1}&=&\sum_{l\ge2}\sum_{k'>l}v_{k'}=\sum_{k'\ge3}(k'-2)v_{k'}=\sum_{k'\ge2}(k'-2)v_{k'}\nonumber\\
&=&\sum_{k'\ge2}k'v_{k'}-2\sum_{k'\ge2}v_{k'}=L-2V,
\eeqa
and so the term is square brackets in \eref{Pmmuzeroeqn2} is identically 0.

This result shows that $P_{m}(\omega,\mu=0)=0$ and hence that $\rho_{m}(t,\mu=0)=0$ for all $m>0$, for both symmetry classes.  This is consistent with the fact that the survival probability should be identically 1 for a closed system ($\mu=0$).

\subsection{Continuity Equation}\label{recursionconteqn}

Building on this, we are now able to treat the full continuity equation in the Fourier space.  Again this will require re-expressing the contribution from the third case, and we also need to sum over different vectors $\v$ with the same value of $L-V=m$.  If we define
\beq
\hat{N}(\v,M)=\frac{(-1)^{V}}{L}\frac{\prod_{\alpha}(M-2\pi\rmi\omega l_{\alpha})}{(M-2\pi\rmi\omega)^{L}}N(\v),
\eeq
then the first two contributions to the transformed survival probability, see equations \eref{PvAeqn} and \eref{PvBeqn}, can be written as 
\beqa
P_{m,\mathrm{A}}(\omega)=&\sum_{\v}^{L-V=m}\frac{L}{(M-2\pi\rmi\omega)}\hat{N}(\v,M) \\
P_{m,\mathrm{B}}(\omega)=&-\sum_{\v}^{L-V=m}\sum_{l}\frac{2lv_{l}}{(M-2\pi\rmi\omega l)} \hat{N}(\v,M),
\eeqa
where we have replaced the sum over $\alpha$ by a sum over the components of the vector $\v$.  To simplify the third contribution from \eref{PvCeqn} we use a version of \eref{twoencorbeqiveqn2}, modified for this situation to include the extra factors
\beqa
&\frac{(-1)^{V}\N_{k,l}(\v)}{(M-2\pi\rmi\omega k)(M-2\pi\rmi\omega l)}\frac{\prod_{\alpha}(M-2\pi\rmi\omega l_{\alpha})}{(M-2\pi\rmi\omega)^{L-1}} \nonumber\\
&= -\frac{(k+l-1)v_{k+l-1}^{[k,l\to k+l-1]}}{(M-2\pi\rmi\omega(k+l-1))}\hat{N}(\v^{[k,l\to k+l-1]},M).
\eeqa
We can then rewrite the sum over the dummy vector $\v'=\v^{[k,l\to k+l-1]}$ as a sum over $\v$ and obtain
\beq
P_{m,\mathrm{C}}(\omega)=-\sum_{\v}^{L-V=m}\sum_{k,l}\frac{(k+l-1)v_{k+l-1}}{(M-2\pi\rmi\omega(k+l-1))}\hat{N}(\v,M).
\eeq

We return to the continuity equation for the off-diagonal terms ($J_{m}(t)+\partial\rho_{m}(t)/\partial t=0$) in the Fourier space \eref{fourierconteq}.  To ensure that the continuity equation holds, we have to check that
\beq \label{contftcompeqn}
\fl M\left[P_{m,\mathrm{A}}(\omega)+\frac{P_{m,\mathrm{B}}(\omega)}{2}\right]-(2\pi\rmi\omega)\left[P_{m,\mathrm{A}}(\omega)+P_{m,\mathrm{B}}(\omega)+P_{m,\mathrm{C}}(\omega)\right]=0.
\eeq
Writing the left hand side in terms of the sum over vectors, we have to evaluate
\beq \label{contfteqn}
\fl \sum_{\v}^{L-V=m}\left[L-\sum_{l}\frac{lv_{l}(M-4\pi\rmi\omega)}{(M-2\pi\rmi\omega l)}+\sum_{l}\sum_{k'>l}\frac{2\pi\rmi\omega k'v_{k'}}{(M-2\pi\rmi\omega k')}\right]\hat{N}(\v,M),
\eeq
where $k'=k+l-1$.  Following similar reasoning to \eref{compsumeqn} we can simplify the double sum inside the square brackets as follows
\beq
\sum_{l}\sum_{k'>l}\frac{2\pi\rmi\omega k'v_{k'}}{(M-2\pi\rmi\omega k')}=\sum_{l}\frac{2\pi\rmi\omega (l-2)lv_{l}}{(M-2\pi\rmi\omega l)},
\eeq
so that \eref{contfteqn} becomes
\beqa 
& & \sum_{\v}^{L-V=m}\left[L-\sum_{l}\frac{lv_{l}(M-4\pi\rmi\omega-2\pi\rmi\omega(l-2))}{(M-2\pi\rmi\omega l)}\right]\hat{N}(\v,M) \nonumber\\
& & = \sum_{\v}^{L-V=m}\left[L-\sum_{l}lv_{l}\right]\hat{N}(\v,M) = 0,
\eeqa
since $\sum_{l}lv_{l}=L$.  This verifies equation \eref{contftcompeqn} and shows that the semiclassical expansion satisfies the continuity equation for all $m>0$.  For the remaining diagonal terms (which can be thought of as corresponding to $m=0$) this can be verified directly.

Both of these proofs rely on being able to re-express the contribution from the third case in terms of a sum over vectors which has a similar form to the other two cases.  This relation is then responsible for the fact that we can shift from the survival probability to the current density via the continuity equation and remove the possibility of having trajectories from case C (0ll).

\section{Implications for transport}\label{transport}

We have seen how by differentiating $\rho(t)$, with respect to time, we obtain $-J(t)$ in line with the continuity equation.  For $\rho(t)$ we have a picture involving trajectories that start and end inside the cavity, and we have three cases to consider.  When we differentiate $\rho(t)$, and shift to $J(t)$, we arrive at a picture in terms of trajectories that start inside the cavity but end in the lead, effectively removing the third case (0ll) (and halving the contribution of 1ll).  The next step is to repeat this process and differentiate again with respect to time.  This leads to the more usual transport picture involving trajectories that start and end in the leads, where we can only have case A (2ll).

We can consider the conservation of the current density, which has its own continuity equation. For systems with constant potentials, like billiards, this continuity equation has the simple form $\partial \j(\r,t)/\partial t+\nabla f(\r,t)=0$, where $f(\r,t)$ is the second (antisymmetric spatial) derivative of the local density
\beq \label{feqn}
\fl f(\r,t)=\frac{\hbar^{2} }{(2m\rmi)^2}[\psi^*(\r,t)\mathbf{\nabla}^{2} \psi
(\r,t)-2\mathbf{\nabla} \psi^*(\r,t)\mathbf{\nabla}\psi (\r,t)+\psi(\r,t)\mathbf{\nabla}^{2} \psi^*(\r,t)].
\eeq
To obtain a semiclassical approximation for this quantity, we express the wavefunction in terms of the semiclassical propagator using equations \eref{psiteqn} and \eref{smcpropeqn} and follow the same steps as in section~\ref{semicurrdens}  
\beq \label{ftrajeqn}
\fl f^{\rm sc}(\r,t)=\frac{1}{4m^2\pi^2 \hbar^2}\int \rmd \r_{\o}\sum_{\gamma,\gamma'(\r_{\o}\to \r,t)}D_{\gamma}D_{\gamma'}^*\rme^{\frac{\rmi}{\hbar}(S_{\gamma}-S_{\gamma'})}\rho_{\mathrm{W}}(\r_{\o},\p_{\gamma\gamma'}^{\o})\left(\p_{\gamma\gamma'}^{\f}\right)^{2}.
\eeq
We are interested in the integrated version of this quantity $F(t)=\int\rmd\r\:\nabla^2 f(\r,t)$, which can be expressed in terms of trajectories starting and ending at the lead cross-section.  Instead of using \eref{ftrajeqn}, which involves trajectories which start and end inside the cavity, we first return to \eref{feqn} and replace all derivatives with respect to $\mathbf{r}$, originating from the derivative with respect to time of the current density, by derivatives with respect to $\mathbf{r}_{\o}$, which is defined before \eref{jtrajeqn2}, and neglecting changes in the amplitudes. This replacement is possible in the case of energy-conserving dynamics.  We can then rewrite the resulting expression as a divergence with respect to $\mathbf{r}$ and $\mathbf{r}_{\o}$ in a similar way as it is done with respect to $\mathbf{r}$ to obtain the continuity equation itself.  Upon applying Gauss' theorem for transforming the integrals with respect to $\mathbf{r}$ and $\mathbf{r}_{\o}$ to surface integrals, and again supposing that the initial wave function has a well defined energy, we arrive at the semiclassical expression
\beq \label{Ftrajtempeqn}
\fl F(t) \approx \frac{1}{4m^2\pi^2 \hbar^2}\int_{S} \rmd x\rmd x' \sum_{\gamma,\gamma'(x\to x',t)}D_{\gamma}D_{\gamma'}^*\rme^{\frac{\rmi}{\hbar}(S_{\gamma}-S_{\gamma'})}p_{x,\gamma\gamma'}p_{x',\gamma\gamma'},
\eeq
which is expressed as a sum over trajectories travelling from the lead to itself.  We note that this result also gives the main semiclassical contribution for more general chaotic systems, and not only those with constant potentials. We can extend the connection to transport by projecting onto the channel basis
\beq \label{Ftrajeqn}
F(t) \approx \frac{1}{T_{\mathrm{H}}^{2}}\sum_{a,b} \sum_{\gamma,\gamma'(a\to b,t)}\tilde{D}_{\gamma}\tilde{D}_{\gamma'}^* \rme^{\frac{\rmi}{\hbar}(S_{\gamma}-S_{\gamma'})},
\eeq
where the sum over $a$ and $b$ is over the channels in the lead and then we sum over trajectories connecting these channels. The projection onto the channel basis changes the form of the stability amplitudes $\tilde{D}_{\gamma}$ slightly, though they can still be treated using open sum rules \cite{rs02}.  This form of $F(t)$ is the Fourier transform of a correlation function of scattering matrix elements.

Because encounters cannot occur in the leads we only have a single case, corresponding to 2ll, and we can perform the sum over correlated trajectories using the open sum rule and an auxiliary weight function as before \cite{heusler06,muller07}.  However we have an additional contribution for systems with time reversal symmetry when the start and end channels coincide ($a=b$).  Then we can also compare the trajectory $\gamma$ with the time reversal of its partner $\overline{\gamma'}$ and we obtain a factor of 2 for this channel combination.  This extra possibility corresponds to coherent backscattering and must be considered more carefully when Ehrenfest time effects are important.  The sum over channel combinations therefore gives a factor of $M(M+\kappa-1)$, where $\kappa=1$ for systems without time reversal symmetry and $\kappa=2$ for systems with time reversal symmetry.  The diagonal contribution is
\beq
F^{\mathrm{diag}}(t)=\mu^{2}\left(1+\frac{\kappa-1}{M}\right)\rme^{-\mu t},
\eeq
while we can simply express the contribution of trajectories described by a vector $\v$ as
\beq \label{Fvconteqn}
F_{\v}(t)=\mu\left(1+\frac{\kappa-1}{M}\right)J_{\v,\mathrm{A}}(t).
\eeq

We again shift to the Fourier space, where because of \eref{Fvconteqn}, the integrated continuity equation
\beq \label{jconteqn}
\frac{\partial}{\partial t}J(t)+F(t)=0,
\eeq
becomes, for the off-diagonal terms
\beq \label{Jcontftcompeqn}
(M+\kappa-1)\J_{\mathrm{A}}(\omega)-(2\pi\rmi\omega)\left[\J_{\mathrm{A}}(\omega)+\J_{\mathrm{B}}(\omega)\right]=-\frac{\mu(\kappa-1)}{(M-2\pi\rmi\omega)},
\eeq
where the term on the right is what is leftover from the diagonal approximation for the orthogonal case.  Rewriting the left hand side in terms of a sum over vectors we have to see if the following holds (dividing through by $\mu$)
\beq \label{Jcontfteqn}
\fl \sum_{m}\sum_{\v}^{L-V=m}\left[L+\sum_{l}\frac{lv_{l}(2\pi\rmi\omega)}{(M-2\pi\rmi\omega l)}+\frac{(\kappa-1)L}{(M-2\pi\rmi\omega)}\right]\hat{N}(\v,M)=-\frac{\kappa-1}{(M-2\pi\rmi\omega)}.
\eeq
Focusing on the left hand side, we recall that $L=\sum_{l} lv_l$ and so we can rewrite the first term in square brackets as 
\beq 
L=\sum_l \frac{lv_{l}(M-2\pi\rmi\omega l)}{(M-2\pi\rmi\omega l)},
\eeq
so that we can combine it with the second term.  We also separate the third term to rewrite the left hand side of \eref{Jcontfteqn} as
\beqa \label{Jcontfteqn2}
\sum_{m}\sum_{\v}^{L-V=m}\sum_{l}\frac{lv_{l}(M-2\pi\rmi\omega(l-1))}{(M-2\pi\rmi\omega l)}\hat{N}(\v,M) \\ \nonumber
+(\kappa-1)\sum_{m}\sum_{\v}^{L-V=m}\frac{L}{(M-2\pi\rmi\omega)}\hat{N}(\v,M).
\eeqa
For the unitary case, the second line vanishes, and because of the result \eref{unitrecresulteqn} in \ref{morerecursion}, we can see that the sum in \eref{Jcontfteqn2} is identically zero for each $m$ and so the continuity equation is satisfied.  For the orthogonal case, using the result \eref{orthrecresulteqn} in \ref{morerecursion}, we can see that the terms in the sum in the first line for $m=k$ cancel with the terms in the sum in the second line where $m=k-1$.  The only term remaining when we sum over all $m$ is therefore the term from the first line where $m=1$.  This corresponds to a vector with a single 2-encounter for which we can easily evaluate
\beq 
\sum_{\v}^{L-V=1}\sum_{l}\frac{lv_{l}(M-2\pi\rmi\omega(l-1))}{(M-2\pi\rmi\omega l)}\hat{N}(\v,M) = -\frac{1}{(M-2\pi\rmi\omega)}.
\eeq
as $N(\v)=1$.  This term cancels exactly with the remaining term from the diagonal approximation, and we have verified \eref{Jcontfteqn} and hence \eref{jconteqn} for both symmetry classes.

\section{Conclusions}\label{conclusions}

The initial result of this article was the calculation of a semiclassical expansion of the integrated current density.  The semiclassical approximation for this quantity involves correlated trajectory pairs which start at some point inside the system and end up escaping through the lead.  The expansion was calculated by working along the lines of \cite{muller07}, but because the trajectories start inside the system we also need to include the contribution of `one-leg-loop' diagrams introduced in \cite{waltner08}.  These, and their extensions, were explored in detail in \cite{gutierrez08} for the survival probability itself, which semiclassically involves trajectories that both start and end in the bulk.  Because of this, there are three cases that need to be considered for the survival probability rather than the two cases which exist for the current density, but both quantities can be connected through the continuity equation.  This connection, however, does not trivially hold semiclassically, so we first showed directly that our calculation for the current density matches up with that for the survival probability in \cite{gutierrez08} which in turn was shown to agree with the supersymmetric random matrix results in \cite{ss97,ss03}.  We note that our methods can also be applied directly to the quantities in the continuity equation \eref{QCE}, though we concentrated on the integrated version \eref{intconteqn} here because of the interesting combinatorics.

We then proceeded to show that, within the semiclassical approximation, this continuity equation is satisfied to all orders.  This proof involved the recursion relation arguments presented in section~\ref{recursion} and is the main result of this article.  This all hinges on our ability to re-express the contribution from the third case, where both the start and end point are inside encounters and which is only present for the survival probability, in terms of contributions related to the other two cases.  The unitarity expressed by the continuity equation is therefore reflected in the combinatorial results derived from considering how many valid trajectory structures can be created by merging together encounters in more complicated trajectories.  Moreover, we can expect this re-expression to lie behind any semiclassical situation where we move from a picture involving trajectories connecting points in the bulk to a picture in terms of trajectories connecting points in the lead. Interestingly, continuity can be thought of as a manifestation of the gauge (phase) invariance of quantum mechanics.  Our proof of the continuity equation shows that this gauge symmetry would also be satisfied semiclassically, but the connection raises the intriguing question of whether treating the gauge invariance directly could lead to a simpler semiclassical proof and also shed more light on the combinatorial structure underlying this work.

An important point, however, is the generality of the continuity equation.  We have shown that it holds semiclassically for chaotic systems in the regime of times shorter than the Heisenberg time, and where Ehrenfest time effects can be ignored.  Moving directly beyond the Heisenberg time is probably the greater challenge, but Ehrenfest time effects have provoked much interest recently, especially due to their appearance in real physical systems.  They have been treated semiclassically \cite{rb06,jw06,br06}, only for low orders (which are the most interesting physically), but this has yet to be generalised.  One thing that we do know is that additional diagrams play a role with finite Ehrenfest time, and these are not included in the formalism used in this article (as they give no contribution in the limit we were working in).  It is therefore an interesting challenge to generalise the extra lower order diagrams and completely extend the semiclassical treatment to the Ehrenfest time regime.  Of course, the continuity equation holds in all regimes and provides the perfect playing ground to explore all the trajectory structures that contribute when Ehrenfest time effects are important, as well as requiring that `one-leg-loops' and their generalisations are included.  As well as Ehrenfest time effects, an obvious extension of the work presented here would be to include tunnel barriers at the lead.  The possible back-reflection when trajectories try to escape leads to additional possible diagrams, as well as to other changes in the system \cite{whitney07}.

The final result of this article was that the current density is itself related, via a continuity equation, to a transport quantity that is expressed in terms of trajectories that start and end in the lead.  This result was also proved, in our semiclassical regime, using more recursion relations derived from \cite{muller05, muller05b}.  More importantly, it shows that we can, in a two-step process, move from a picture involving pairs of trajectories connecting points in the bulk to the more usual transport trajectories which connect points in the lead.  From the survival probability, this involves differentiating twice (in line with the continuity equations) and taking a Fourier transform to arrive at a correlation function of the scattering matrix, which itself is closely linked to the conductance.

\ack{The authors wish to thank the DFG for funding under GRK 638 and FOR 760.}

\appendix \section{More recursion relations} \label{morerecursion}

To prove our second continuity equation we return to the combinatorics of the number of periodic orbit structures described by a vector $\v$ \cite{muller05, muller05b}, and we start with the unitary case.  By considering the number of ways a 2-encounter could merge with a $k$-encounter (to form a $(k+1)$-encounter) by shrinking the links connecting the 2-encounters of the structure, they arrived at the relation
\beq
\frac{2v_{2}}{L}N(\v)=\sum_{k\geq2}\frac{(k+1)(v_{k+1}+1)}{L-1}N(\v^{[2,k\to k+1]}),
\eeq
where $\v^{[2,k\to k+1]}$ is the vector formed from $\v$ by combining a 2-encounter and a $k$-encounter to form a $(k+1)$-encounter, so that $v_2$ and $v_k$ are reduced by one, while $v_{k+1}$ is increased by one.  We want to turn this relation into a version for open systems involving the extra factors we have, in a similar way as was done for parametric correlations in \cite{nagao07}.  Including the extra terms and rearranging, we arrive at the following
\beqa \label{unitopenreceqn}
\frac{2v_{2}(M-2\pi\rmi\omega)}{(M-4\pi\rmi\omega)}\hat{N}(\v,M) \\ \nonumber
+ \sum_{k\geq2}\frac{(k+1)v_{k+1}^{[2,k\to k+1]}(M-2\pi\rmi\omega k)}{(M-2\pi\rmi\omega(k+1))}\hat{N}(\v^{[2,k\to k+1]},M)=0,
\eeqa
where $v_{k+1}^{[2,k\to k+1]}=v_{k+1}+1$.  Because this is identically zero, if we sum over all vectors with a common value of $L-V=m$ the result is still zero
\beqa \label{unitopensumeqn}
\sum_{\v}^{L-V=m} \left[\frac{2v_{2}(M-2\pi\rmi\omega)}{(M-4\pi\rmi\omega)}\hat{N}(\v,M) \right. \\ \nonumber
\left. + \sum_{k\geq2}\frac{(k+1)v_{k+1}^{[2,k\to k+1]}(M-2\pi\rmi\omega k)}{(M-2\pi\rmi\omega(k+1))}\hat{N}(\v^{[2,k\to k+1]},M)\right]=0.
\eeqa
As combining a 2-encounter and a $k$-encounter reduces both $L$ and $V$ by one, the resulting vector $\v^{[2,k\to k+1]}$ has the same value of $L-V=m$ as $\v$.  The important step is then to express the sum over the resulting vectors as
\beq \label{resumeqn}
\sum_{\v}^{L-V=m} v_{k+1}^{[2,k\to k+1]}\hat{N}(\v^{[2,k\to k+1]},M)= \sum_{\v'}^{L-V=m} v'_{k+1}\hat{N}(\v',M).
\eeq
If we then identify the dummy vector $\v'$ with $\v$ in our original sum, we can rewrite \eref{unitopensumeqn} as
\beq 
\fl \sum_{\v}^{L-V=m}\left[\frac{2v_{2}(M-2\pi\rmi\omega)}{(M-4\pi\rmi\omega)}+\sum_{l\geq3}\frac{lv_{l}(M-2\pi\rmi\omega (l-1))}{(M-2\pi\rmi\omega l)}\right]\hat{N}(\v,M)=0,
\eeq
where $l=k+1$.  The first term can be included as the $l=2$ term in the sum over $l$, so the result reduces to
\beq \label{unitrecresulteqn}
\sum_{\v}^{L-V=m}\sum_{l\geq2}\frac{lv_{l}(M-2\pi\rmi\omega (l-1))}{(M-2\pi\rmi\omega l)}\hat{N}(\v,M)=0.
\eeq

For the orthogonal case, we can also create a valid periodic orbit structure if we shrink a link that connects a 2-encounter to itself so that the 2-encounter disappears.  This means that there is an extra term in the recursion relation.  Recasting the relation from \cite{muller05, muller05b} into a form we require for our situation, we obtain
\beqa \label{orthopenreceqn}
\fl \frac{2v_{2}(M-2\pi\rmi\omega)}{(M-4\pi\rmi\omega)}\hat{N}(\v,M)+\sum_{k\geq2}\frac{(k+1)v_{k+1}^{[2,k\to k+1]}(M-2\pi\rmi\omega k)}{(M-2\pi\rmi\omega(k+1))}\hat{N}(\v^{[2,k\to k+1]},M) \nonumber \\ 
+\frac{L(\v^{[2\to]})}{(M-2\pi\rmi\omega)}\hat{N}(\v^{[2\to]},M)=0,
\eeqa
where $\v^{[2\to]}$ is the vector formed from $\v$ by removing a 2-encounter and $L(\v^{[2\to]})$ is the number of links that the new structure has.  As well as the resummation in \eref{resumeqn} we can also express
\beq \label{resumeqn2}
\sum_{\v}^{L-V=m} L(\v^{[2\to]})\hat{N}(\v^{[2\to]},M)= \sum_{\v'}^{L-V=m-1} L\hat{N}(\v',M),
\eeq
but now, because removing a 2-encounter reduces $L$ by two and $V$ by only one, the value of $L-V$ of our new summation variable $\v'$ is one less than that of $\v$.  Using these resummations, and the fact that when we sum the relation \eref{orthopenreceqn} over all vectors $\v$ with the same value of $L-V=m$ the sum is still zero, we obtain a result of
\beqa \label{orthrecresulteqn}
\sum_{\v}^{L-V=m}\sum_{l\geq2}\frac{lv_{l}(M-2\pi\rmi\omega (l-1))}{(M-2\pi\rmi\omega l)}\hat{N}(\v,M) \\ \nonumber
+\sum_{\v}^{L-V=m-1}\frac{L}{(M-2\pi\rmi\omega)}\hat{N}(\v,M)=0.
\eeqa
This result, along with \eref{unitrecresulteqn}, allows us to prove \eref{jconteqn}, the second of our continuity equations for both symmetry classes.

\end{document}